# Atomic dynamics and local structural disorder during ultrafast melting of polycrystalline Pd

Adam Olczak[1*], Ryszard Sobierajski[2], Przemyslaw Dziegielewski[1], Salman Ali Kahn[1], Zuzanna Kostera[1], Kirill P. Migdal[3], Igor Milov[4,5,6], Klaus Sokolowski-Tinten[7], Peter Zalden[8], Vasily V. Zhakhovsky[3], and Jerzy Antonowicz[1*]

[1]*Faculty of Physics, Warsaw University of Technology, Koszykowa 75, 00-662 Warsaw, Poland.*

[2]*Institute of Physics of the Polish Academy of Sciences, al. Lotników 32/46, 02-668 Warsaw, Poland.*

[3]*Joint Institute for High Temperatures, Russian Academy of Sciences, 13/2 Izhorskaya st., Moscow 125412, Russia.*

[4]*Advanced Research Center for Nanolithography (ARCNL), Science Park 106, 1098 XG Amsterdam, The Netherlands.*

[5]*Deutsches Elektronen Synchrotron DESY, Center for Free Electron Laser Science CFEL, Notkestr. 85, 22607 Hamburg, Germany.*

[6]*Industrial Focus Group XUV Optics, MESA+ Institute for Nanotechnology, University of Twente, Drienerlolaan 5, 7522 NB Enschede, The Netherlands.*

[7]*Center for Nanointegration Duisburg-Essen (CENIDE), University of Duisburg-Essen, Lotharstrasse 1, 47048 Duisburg, Germany.*

[8]*European XFEL, Holzkoppel 4, 22869 Schenefeld, Germany.*

*Corresponding authors:

Adam Olczak, Faculty of Physics, Warsaw University of Technology, ul. Koszykowa 75, 00-662 Warsaw. Poland, e-mail: jerzy.antonowicz@pw.edu.pl

Jerzy Antonowicz, Faculty of Physics, Warsaw University of Technology, ul. Koszykowa 75, 00-662 Warsaw. Poland, e-mail: jerzy.antonowicz@pw.edu.pl

**Abstract**

The primary distinction between solid and liquid phases is mechanical rigidity, with liquids having a disordered atomic structure that allows flow. While melting is a common phase transition, its microscopic mechanisms still remain unclear. This study uses molecular dynamics simulations to investigate ultrafast melting in polycrystalline palladium, focusing on the relationship between atomic dynamics quantified by the root-mean-squared displacement (RMSD) and local structural disorder characterized by the deviation from centrosymmetry. In the crystal bulk, melting is preceded by a gradual rise in the RMSD and local disorder. As the Lindemann limit for the RMSD is approached, the increasing concentration of lattice defects is manifested by a discontinuous rise in disorder. On melting, the rise is followed by a rapid increase in displacement, indicative of atomic flow. In contrast, the grain boundaries undergo melting through a continuous increase of both the displacement and the disorder, resembling a glass transition on heating.

Fluidity and an irregular atomic arrangement are the hallmarks of the liquid state, making it fundamentally distinct from a crystal. On melting, a first-order phase transition, the viscosity and degree of atomic order vary abruptly. While studied over a century, melting, similar to other first-order transitions, lacks a deep understanding [1]. It might appear surprising that up till now, a physical model of melting involving both the solid and the liquid phase is missing. Instead, several melting criteria have been proposed in the literature, predicting critical values for various macro- and microscopic crystal parameters. Once reached, this results in its transition to the liquid state. The macroscopic criterion introduced by Born [2] refers to the crystal's response to shear stress, i.e., its mechanical rigidity. Its core is the "rigidity catastrophe" – the vanishing of the crystal shear modulus on melting. From the microscopic perspective, the melting transition can also be considered a disorder phenomenon [3]. In this view, the crystal melts when its atoms oscillate about their lattice sites so strongly that "direct contact" of neighboring atoms occurs. As a result, the crystalline lattice breaks down, and atoms start to move relative to each other, making the system mobile and non-rigid. This idea expressed by Frank [3] as a crystal "shaking itself to pieces" is the main assumption of Lindemann's theory of fusion [4]–[6]. According to this approach, melting occurs when the root-mean-square displacement (RMSD) of atoms oscillating about their equilibrium lattice positions reaches a critical fraction of the interatomic distance. The fraction, known as the Lindemann parameter ($\delta_L$) varies with crystal structure, but its empirical value for many simple solids is typically in the range of 0.12-0.13 [1]. In real crystals, the amplitude of vibrations of the surface and interface atoms is larger than in the interior. Therefore, unless free surfaces are eliminated [7] or ultrahigh heating rates are involved [8], melting starts preferentially from the crystal surfaces or grain boundaries (GBs) [9]–[12], limiting the possibility of crystal superheating. Jin *et al.* [13] found $\delta_L \cong 0.22$ at the limit of superheating in the computer-simulated infinite (involving 3D periodic boundary conditions) Lennard-Jones monocrystal.

The same value was observed for surface melting at the equilibrium melting point [14], indicating an underlying connection between bulk (homogenous) and surface-mediated (heterogenous) melting.

Apart from the Lindemann model referring to the dynamics of atomic thermal vibrations, several models correlating the melting of a crystal with its point and line defects concentration were proposed [15]. A melting model based on vacancies initially introduced by Frenkel [16] was later developed by Górecki [17], who evidenced that many metals melt when the equilibrium concentration of vacancies reaches a critical value of 0.37% and an additional 10% of vacancies is produced at the expense of the latent heat of fusion. Recently, Mo *et al.* [18] demonstrated by ultrafast electron diffraction that laser-induced melting of radiation-damaged tungsten starts below its melting point. The effect was attributed to the excess population of vacancies, supporting the idea that vacancies are responsible for the melting of metals. Different authors pointed out that the melting crystal simultaneously satisfies different melting criteria. In particular, the vibrational and elastic instabilities were found to coincide at the melting point [13]. On the other hand, dislocation formation requires net atomic displacements, which compares well with that predicted by the Lindemann criterion [19]. The relevance of different criteria for predicting the melting temperature suggests a deep link between the dynamics of atomic motions and the degree of structural disorder. This work focuses on certain aspects of the latter relationship in of a molecular dynamics (MD) simulation of ultrafast, laser-driven melting of polycrystalline Pd. Strong optical excitation creates highly non-equilibrium conditions that kinetically hinder heterogeneous melting and enhance the contribution of homogeneous melting. This reveals a correlation between the dynamics and disorder in the presence of two competing melting mechanisms.

In our study, we employed large-scale MD simulations coupled to a two-temperature model description of the electron dynamics and the electron-lattice energy transfer. In the simulations, the Pd system (20×23×30 nm³) was sandwiched between two 300 nm thick layers (in the y-z plane) of an effective amorphous material representing $Si_3N_4$ substrate and $SiO_2$ cap. Periodic boundary conditions were applied in the y and z directions, simulating infinite layers, while the system was free to expand along the x-axis to model vacuum surrounding. The details of the MD simulations are described in our recent work, which provides the experimental validation of the current numerical predictions [20]. To quantify the local atomic disorder, we employed the modified centrosymmetry parameter originally introduced by Kelchner *et al.* [21] to characterize lattice defects in fcc metals. The modification involves normalizing the parameter to the number of nearest neighbors (NNs) and the NNs interatomic distance. The normalized centrosymmetry parameter is defined as $\text{nCSP} = \frac{1}{Nd^2}\sum_{i=1}^{N/2}|\mathbf{r}_i + \mathbf{r}_{i+N/2}|^2$, where $N$ is the number of NNs, $d$ is the NN interatomic distance, and $\mathbf{r}_i$, $\mathbf{r}_{i+N/2}$ are two vectors corresponding to pairs of opposite NNs. The parameter defined above can be interpreted as a measure of deviation from perfect centrosymmetry within the crystal lattice, expressed as a fraction of $d$. The dynamics of atomic motions is characterized by the magnitude of their displacement $u$ calculated with respect to their original positions. Due to the thermal expansion of the MD system in the x-direction, the displacement representing the cumulative effect of oscillatory (directed) and statistical (diffusive) motion of a given atom was calculated only for the y-z plane according to the formula $u = \sqrt{(y-y_0)^2 + (z-z_0)^2}$ where, y and z are the instantaneous and $y_0$, $z_0$ are the initial coordinates of an atom.

Figure 1 presents the snapshots of the MD system cross-section for selected times $t$ elapsed since laser excitation (0.5 ps, 2,5 ps, 5 ps, 10 ps) and deposited energy densities $E$ (0.48 MJ/kg, 0.70 MJ/kg, 1.4 MJ/kg). The lowest $E$ corresponds to the melting onset; the intermediate one illustrates a partial transition into the liquid state, and the highest energy density causes complete melting. The two colouring schemes represent the nCSP and the $u$ of individual atoms. The results in Fig. 1 demonstrate two pathways towards melting the polycrystalline metal, namely (i) a heterogeneous increase of the nCSP in the initially disordered GB regions, and (ii) a homogeneous increase of the nCSP inside the bulk of the grains. The two competing processes progress parallel, yet their contribution varies with absorbed energy density. At low *E*, the local increase of the nCSP in the GBs is more pronounced. However, at strong excitation, the homogeneous increase of the nCSP dominates. The actual distribution of the liquid phase is more clearly visualized when the nCSP snapshots are analyzed in parallel to those of $u$. It is important to remember that the displacement of atoms is due to both atomic oscillations about the equilibrium positions in the crystalline lattice and diffusion (i.e. motion beyond the NN distance). Since atomic diffusion is significantly faster in the liquid than in the solid, the high-$u$ atoms are considered to belong to the molten regions of the system. It can be seen in Fig. 1 that the spatial distribution of the high-$u$ atoms in the partially molten state is non-uniform at low (0.48 MJ/kg) and intermediate (0.70 MJ/kg) energy densities. The distribution of high-$u$ atoms reflects the location of the GBs and correlates with that of the nCSP. At high $E$ (1.4 MJ/kg), $u$, similarly to the nCSP, is enhanced mostly uniformly, indicating the prevalence of the homogenous melting mechanism.

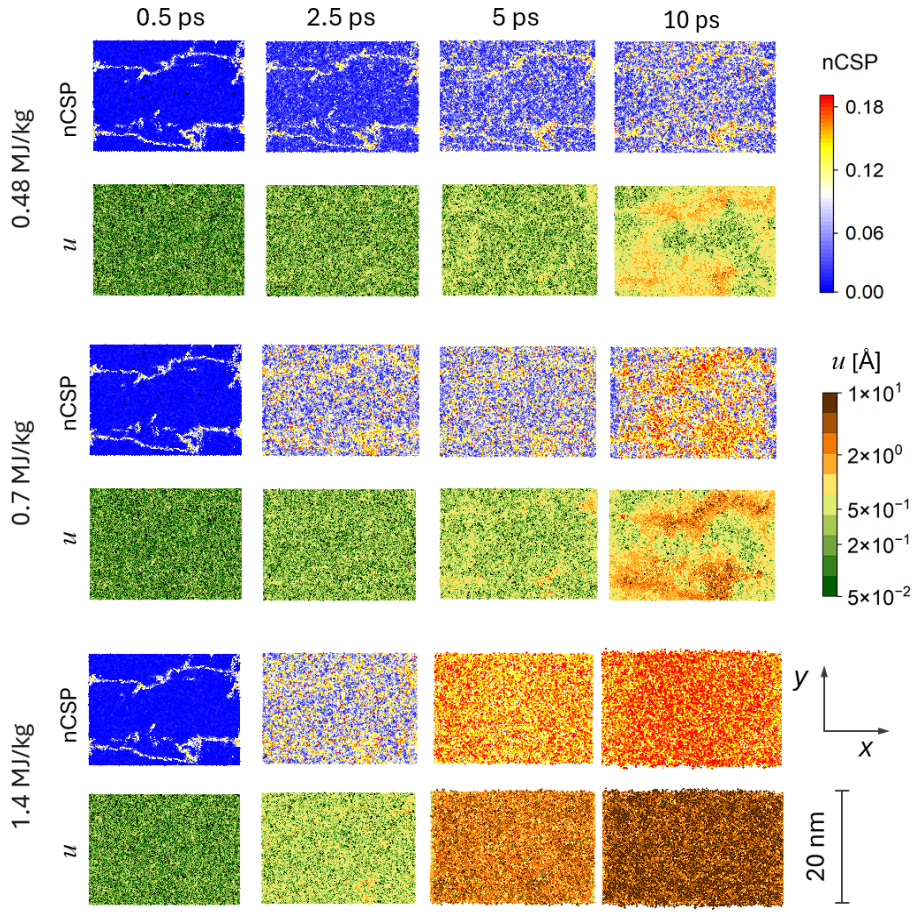

Figure 1. The snapshots of the MD system cross-section for selected times t (0.5 ps, 2.5 ps, 5 ps, 10 ps) and absorbed energy densities E (0.48 MJ/kg, 0.70 MJ/kg, 1.4 MJ/kg). The two colouring schemes denote the normalized centrosymmetry parameter nCSP and the magnitude of displacement u of the individual atoms.

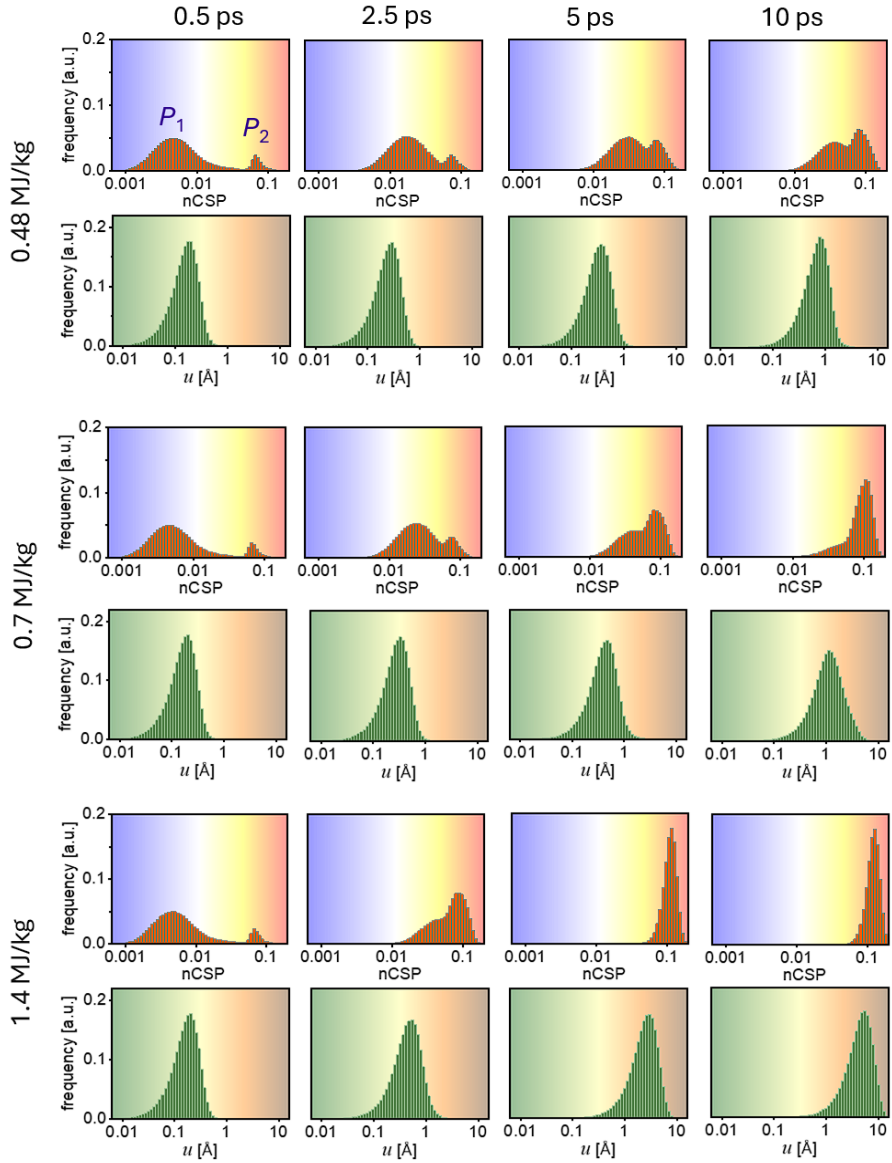

Figure 2. The histograms of nCSP and u plotted for selected times (0.5 ps, 2.5 ps, 5 ps, 10 ps) and absorbed energy densities (0.48 MJ/kg, 0.70 MJ/kg, 1.4 MJ/kg) corresponding to those in Fig. 1. The two distinct nCSP peaks are marked as $P_1$ and $P_2$. The background colors correspond to the color scales of the Fig. 1.

A significant limitation of the presentation depicted in Fig. 1 is the low fraction of visible atoms relative to the overall size of the MD system. To track the global variation of structural disorder and atomic dynamics, we assessed the overall distributions of the nCSP and $u$ at various stagesf the melting process. Fig. 2 presents histograms depicting the distributions of the nCSP and $u$ corresponding to the results shown in Fig. 1. At 0.5 ps after excitation, the nCSP distribution reveals two peaks: a prominent peak centred around 0.005 (further referred to as $P_1$) and a secondary, weaker peak centered around 0.07 ($P_2$). A comparison with Fig. 1 indicates that $P_1$ represents a distribution of local disorder among atoms belonging to the bulk of the grains, whereas $P_2$ corresponds to the small fraction of atoms localized within the GBs. Two primary trends are evident in the temporal evolution of the nCSP shown in Fig. 2 are (i) a gradual shift of the $P_1$ and $P_2$ positions towards higher nCSP values, and (ii) an increase of the number of atoms constituating $P_2$ at the expense of the $P_1$-population. In contrast, the distribution of $u$ shows a single maximum that progressively shifts towards higher displacement values.

The data presented in Fig. 1 and Fig. 2 illustrate the variations of the nCSP and $u$ yet do not explicitly inform on the correlation between the two parameters. Plotting density maps of points representing individual atoms on $u$ vs. nCSP coordinate plane provides a graphical representation of the correlation and its evolution over time. Such a visualization, shown for the intermediate energy density in Figure 3, reveals patterns and trends in the relationship between atomic dynamics and local structural disorder as the system undergoes melting. As demonstrated in Fig. 3, the progress of the solid-to-liquid transition involves the "hopping" of atoms between two clusters of points on the $u$ vs. nCSP coordinate plane: cluster $C_1$ represents the ordered state (corresponding to the peak $P_1$ in Fig. 2) and cluster $C_2$ (corresponding to $P_2$) represents the disordered state. According to Fig. 3, the "hopping" occurs only once a threshold value (estimated around 0.33 Å) of $u$ is exceeded. By determining an average value of $u$ for each cluster, which is equivalent to the root-mean-squared displacement (RMSD), one can normalize the RMSD to the equilibrium interatomic distance in the simulated Pd crystal (2.72 Å). This normalization allows relating the threshold to the Lindemann ratio $\delta_L$, which in the current case yields 0.12, as expected.

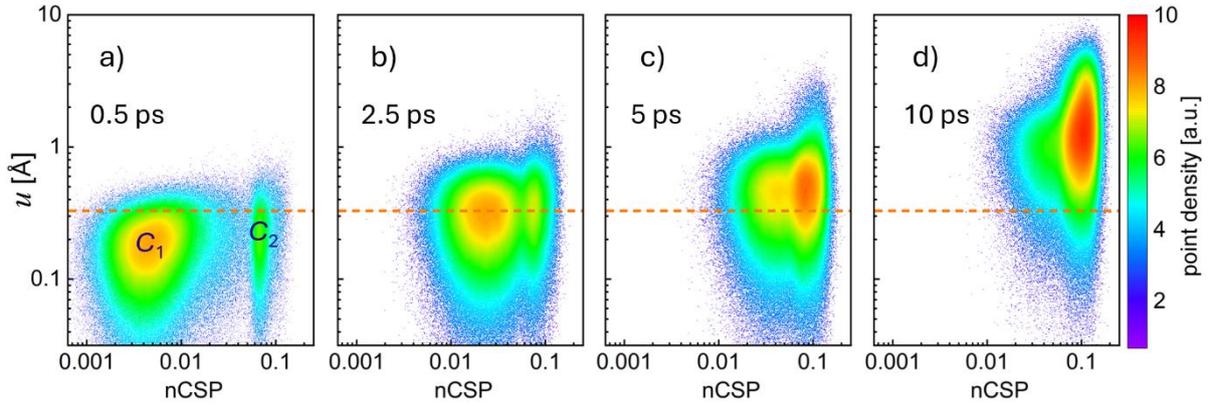

Figure 3. Time-dependent correlation between the $u$ and the nCSP for excitation of 0.7 MJ/kg corresponding to an incomplete melting. The horizontal dashed line marks the value of the Lindemann parameter equal to 0.12. The color coding represents the density of the data points, where each point corresponds to one of the 440 000 atoms (a quarter of the entire MD system). The two clusters of points are denoted as $C_1$ and $C_2$.

The results for the intermediate $E$ shown in Fig. 3 are replotted on the RMSD vs cCSP vs. RMSD coordinate system in Fig. 4 together with the low- and high-$E$ data in a concise way, allowing observation of the temporal evolution of the populations of $C_1$ and $C_2$. At early times, independently of the energy density, the locations of $C_1$ and $C_2$ shift towards higher nCSP and RMSD values. This stage corresponds to an increase in the amplitude of atomic vibrations due to the transfer of energy from the hot electron gas to the lattice and the resulting thermal disordering of the crystal. As soon as the RMSD reaches the Lindemann limit (horizontal dashed line in Fig. 4), the correlation trend for $C_1$ and $C_2$ changes, and the increase of the RMSD becomes more rapid. In parallel, the population of $C_2$ starts to increase at the expense of the intensity of $C_1$. In this time regime, the mobility of atoms significantly increases, and the system undergoes a discontinuous transition in the local atomic environment reflected by an abrupt variation of centrosymmetry. Unless complete melting is achieved, $C_1$ and $C_2$ do not merge but exist as two distinct clusters of points, as evidenced by the vertical gap between $C_1$ and $C_2$ in Fig. 4. Also, as seen in the figure, the $C_1$ cluster persists for RMSD values exceeding the Lindemann range. Close examination of the MD data indicates that the effect is due to the diffusive atomic jumps occurring in the hot crystal and not thermal vibrations (see Supplementary Material). The existence of two distinct pathways for the bulk ($C_1$) and the GB ($C_2$) atoms towards melting are marked by traces of the two clusters. As long as the displacement of oscillating atoms remains below the value corresponding to $\delta_L$, the disorder in the crystal bulk and GBs increases continuously. According to Fig. 4, the GB atoms reach $\delta_L$ slightly sooner than those belonging to the bulk atoms which undergo a discontinuous transition between the ordered ($C_1$) and disordered/defective ($C_2$) states. The location of $C_2$ at the Lindemann limit (nCSP $\cong$ 0.08, RMSD $\cong$ 0.33 Å) marks the onset of melting, manifested by a rapid increase in the displacement and further rise of the local disorder, up to nCSP $\cong$ 0.10.

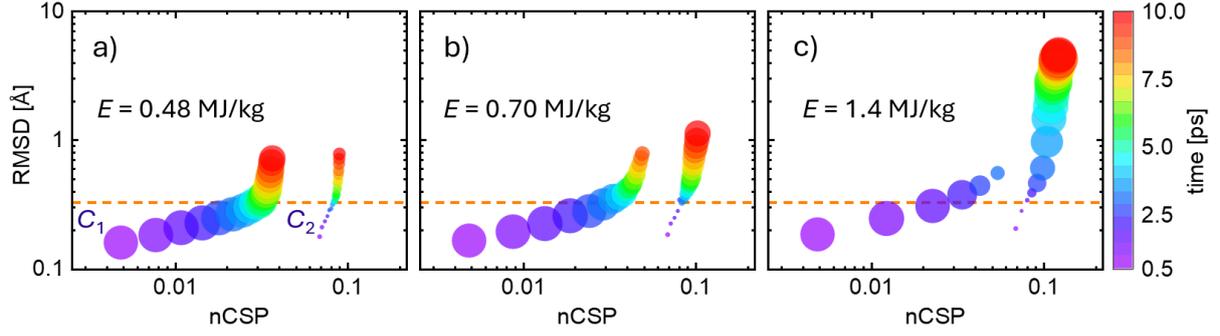

Figure 4. Time-dependent correlation between the atomic RMSD and the nCSP at different deposited energy density values. The horizontal dashed line marks the value of the Lindemann parameter equal to 0.12. The size of the data points represents the height of the point density maxima of clusters $C_1$ and $C_2$ in the $u$ vs. nCSP coordinate plane (see colour coding in Fig. 3).

Notably, the nCSP value of 0.08 is typical for atoms surrounding a single vacancy in the fcc Pd lattice [20].

The observed trace of cluster $C_2$ on the RMSD vs. nCSP coordinate plane of Fig. 4 indicates that the liquid-type local order is similar to that in the disordered GB regions. We argue that for the GB atoms, the temperature increase involves a slight variation in the local atomic environment, which, on melting, leads to a rapid increase in mobility. In this sense, the melting initiated in the GBs resembles the glass transition during heating when the configurationally frozen disordered system transitions to a structurally similar (supercooled) liquid, accompanied by a remarkable increase in mobility. This observation is consistent with the previous hypothesis on structural and dynamic similarities between GBs and glass-forming liquids [10], [22]. In particular, Wolf [10] pointed out that the combined static and dynamic (thermal) disorder of the GBs must reach a certain level for their solid-to-liquid transition to occur, suggesting behavior reminiscent of the Lindemann criterion for melting. Our current results provide further support for this view.

In summary, our study of strongly excited polycrystalline Pd undergoing melting demonstrates an underlying interrelation between the dynamics of atomic motions and the degree of their local structural disorder. This interrelation allows for unifying heterogeneous (GB-mediated) and homogeneous (bulk) melting within a single framework, where atomic displacements satisfying the Lindemann criterion for melting coincide with reaching a characteristic level of disorder in the local atomic environment. According to our results, the RMSD exceeding the Lindemann limit $\delta_L$ is first reached by atoms located in hot GBs and those in the vicinity of lattice point defects of the crystal bulk, such as vacancies. Melting in the crystal bulk is preceded by a gradual rise of atomic displacements and local disorder resulting from the increasing amplitude of atomic thermal oscillations. As melting is approached, the increasing concentration of crystal defects is manifested by a discontinuous rise in the disorder of the defect-neighbouring atoms. The melting transition is manifested by a minor increase in local disorder and a rapid increase in displacement of these atoms, indicative of atomic flow. In contrast, the melting of disordered GB regions involves an increase in the RMSD and a continuous variation in the nCSP, resembling a glass transition. The presented results indicate that the correlation between the local atomic disorder and dynamics of atomic motions is the universal foundation for understanding the pathway of material transition from the solid to the liquid state.

**Acknowledgments**

This work was supported by the National Science Centre, Poland, grant agreement No 2021/43/B/ST5/02480. This research was carried out with the support of the Interdisciplinary Centre for Mathematical and Computational Modelling University of Warsaw (ICM UW) under the computational allocation No G96-1896.

IM gratefully acknowledges financial support from Dutch Research Council (NWO) (Project 'PROMT', Grant Rubicon Science 2021–1 S, file number 019.211EN.026), and the Industrial Partnership Program 'X-tools', project number 741.018.301.

KST acknowledges support by the Deutsche Forschungsgemeinschaft (DFG, German Research Foundation) through the Collaborative Research Centre 1242 (project number 278162697-SFB1242, project C01 Structural Dynamics in Impulsively Excited Nanostructures).

The contributions of V.V.Z. and K.P.M. to this work were carried out as part of a scientific collaboration during the calendar year 2021.